\documentclass[aps,amsmath,amssymb,amsfonts,twocolumn]{revtex4}
\usepackage{epsfig}
\usepackage{graphicx}
\usepackage{subfigure}
\usepackage{dcolumn}
\usepackage{bm}
\usepackage{amsthm}
\usepackage{enumitem}
\usepackage{slashed}
\usepackage{braket}
\usepackage{amsmath}
\usepackage{mathtools}

\makeatletter
\def\l@subsubsection#1#2{}
\makeatother

\newcommand{\del}{\partial}

\newcommand{\bbP}{\mathbb{P}}

\newcommand{\calL}{\mathcal{L}}

\newcommand{\calO}{\mathcal{O}}

\begin{document}

\title{Universal structure in the interactions of massless fields on the lightcone}

\author{Jin Kozaki}
\email{jin.kozaki53@gmail.com}
\affiliation{Okinawa Institute of Science and Technology, 1919-1 Tancha, Onna-son, Okinawa 904-0495, Japan}

\author{Yasha Neiman}
\email{yashula@icloud.com}
\affiliation{Okinawa Institute of Science and Technology, 1919-1 Tancha, Onna-son, Okinawa 904-0495, Japan}

\date{\today}

\begin{abstract}
We consider cubic interactions of massless fields of arbitrary spin in 4 spacetime dimensions, within the lightcone formalism. We extend two key results from flat to (Anti-)de Sitter spacetime. First, we present a simple explicit expression for all the cubic vertices. Second, we prove that the various self-dual/chiral theories, such as Self-Dual Yang-Mills, Self-Dual General Relativity and Chiral Higher-Spin Gravity, are fully consistent with no need for vertices beyond cubic. The key observation behind our results is that all cubic vertices in the lightcone formalism can be expressed as a direct generalization of the ``abelian'' vertices formed by multiplying linearized curvatures. The simple relationship between flat and (Anti-)de Sitter then follows essentially from the conformal invariance of such linearized curvatures. The price for such simplicity is that our vertex expressions are non-local. It is however easy to bring them into a local form, which we also present.
\end{abstract}

\maketitle

\section{Introduction} \label{sec:intro}

In this paper, we discuss the structure of cubic interactions of spinning massless fields in 4d spacetime (restricting to bosons for simplicity). This includes the cubic interactions of Yang-Mills and General Relativity (GR), within themselves, with each other, and with massless matter. In the lightcone formalism, all such fields and interactions can be viewed as pieces within a ``master theory'' of interacting massless fields with all spins: Higher-Spin (HS) Gravity \cite{Vasiliev:1990en,Vasiliev:1995dn,Vasiliev:1999ba}. The latter is, in its own right, an important subject within mathematical physics -- as the 4d holographic dual of 3d vector models \cite{Klebanov:2002ja} and more general Chern-Simons-matter theories \cite{Giombi:2012ms}, as a model of holography in de Sitter space \cite{Anninos:2011ui}, as a cousin of supergravity with its own version of BPS objects \cite{Didenko:2009td,David:2020fea,Lysov:2022zlw}, as a subsector of tensionless strings \cite{Gaberdiel:2015wpo}, and as a dual description of string theory itself \cite{Chang:2012kt}.

Every spin-$s$ degree of freedom with helicity $h=\pm s$ is described in the lightcone formalism by a ``scalar'' field $\Phi_h$, with no gauge redundancy (the fields might carry color indices, which we suppress). Real fields imply a complex-conjugation relation $\Phi_{-h} = \bar\Phi_h$. We work in Poincare coordinates $x^\mu=(t,x,y,z)$. The spacetime metric is either flat $\eta_{\mu\nu}$, de Sitter (dS) $\eta_{\mu\nu}/t^2$, or Anti-de Sitter (AdS) $\eta_{\mu\nu}/z^2$. In all these cases, the free-field quadratic Lagrangian reads simply:
\begin{align}
	\calL_{[2]} = \frac{1}{2}\sum_h\Phi_{-h}\Box\Phi_h \ , \label{eq:free}
\end{align}
where both the d'Alembertian $\Box$ and the implicit integration measure $d^4x$ are defined w.r.t. the flat $\eta_{\mu\nu}$. HS Gravity contains one field of each integer (or even) helicity, while realistic lower-spin theories are restricted to $-2\leq h\leq 2$. 

The allowed cubic vertices and their detailed form are fixed by \emph{consistency at the cubic order} of the spacetime symmetries, which are non-manifest in the lightcone formalism. There's a unique consistent cubic vertex for every triple of helicities with $h_{\text{tot}}\equiv \sum_{i=1}^3 h_i \neq 0$ \cite{Bengtsson:1983pd,Fradkin:1991iy,Metsaev:2018xip}, all of which appear in HS gravity \cite{Ponomarev:2016lrm,Skvortsov:2018jea,Skvortsov:2018uru}; the only other allowed vertex is the purely-scalar $h_i=(0,0,0)$, which we will not consider. Yang-Mills interactions are covered by the vertices with $h_{\text{tot}} = \pm 1$, and gravitational ones -- by those with $h_{\text{tot}} = \pm 2$. It's helpful to separate the cubic vertices into ``chiral'' ones with $h_{\text{tot}}>0$, denoted collectively as $V_{[3]}$, and their ``anti-chiral'' complex conjugates with $h_{\text{tot}}<0$, denoted as $\bar V_{[3]}$. The theory then takes the form:
\begin{align}
	\calL &= \calL_{[2]} - c\,V_{[3]} - \bar c\,\bar V_{[3]} + O(c\bar c) \ ; \label{eq:L_full} \\
	V_{[3]} &= \sum_{h_{\text{tot}}>0} V_{h_1h_2h_3} \ , \label{eq:V_total}
\end{align}
where we introduced an overall complex coupling constant $c$ for the chiral interactions. The $O(c\bar c)$ term denotes quartic and higher-order vertices. The notation \eqref{eq:L_full} expresses a \emph{powerful, non-trivial statement} about massless theories: if we complexify and set $c\neq \bar c = 0$, then the resulting ``self-dual'' or ``chiral'' theory $\calL_{[2]} + cV_{[3]}$ is \emph{consistent to all orders}, without no need for higher vertices. For short, we'll refer to this property as \emph{higher-order consistency}. The ``master'' theory in this class is Chiral HS Gravity \cite{Ponomarev:2016lrm,Skvortsov:2018jea,Skvortsov:2018uru,Sharapov:2022faa,Sharapov:2022awp}, with all helicities $h$ and all vertices $h_{\text{tot}}>0$. Smaller self-contained theories include Self-Dual Yang-Mills ($h=\pm 1$, $h_{\text{tot}}=1$) and Self-Dual GR ($h=\pm 2$, $h_{\text{tot}}=2$), as well as intermediate theories \cite{Ponomarev:2017nrr,Krasnov:2021nsq,Serrani:2025owx} that include HS fields but only vertices with $h_{\text{tot}}=1,2$. Such self-dual/chiral theories are important for describing instantons \cite{Coleman:1978ae,Hawking:1976jb} (Euclidean solutions that govern tunneling), as well as MHV scattering amplitudes, which are the starting point for all amplitude calculations in flat spacetime \cite{Bardeen:1995gk,Chalmers:1996rq,Rosly:1996vr,Mason:2009afn}. Note that for HS Gravity, the chiral theory exists in both flat and (A)dS versions, while the full theory ($c,\bar c$ both nonzero) likely requires a cosmological constant to tame its non-locality.

The vertices $V_{[3]}$ and their higher-order consistency are known in flat spacetime, but only piecemeal in (A)dS. Our aim is to fill this gap. In the process, we'll uncover a universal structure that relates all cubic vertices, both flat and (A)dS, to the ``abelian'' ones constructed from curvature tensors. We build on several previous works. In \cite{Metsaev:1999ui,Metsaev:2003cu}, an AdS lightcone formalism was developed, and the vertices $V_{[3]}$ were found ``in principle'' \cite{Metsaev:2018xip}, as a formal solution to a complicated differential equation. In \cite{Kozaki:2025jrj}, the results of \cite{Metsaev:2018xip} were extended from AdS to de Sitter (where the warp factor $t$ can't be orthogonal to the preferred lightlike direction). For Self-Dual GR and some cousins with $h_{\text{tot}}=1,2$, higher-order consistency in (A)dS can be inferred from the lightcone-gauge results of \cite{Neiman:2023bkq,Neiman:2024vit}, which inspired the present work.

\section{The cubic vertices}

In the lightcone formalism, the price of avoiding gauge redundancy is the loss of manifest Lorentz invariance. We single out a lightlike direction (equivalently, directional derivative) $\del_-$, which defines a left-handed complex plane $\del_-\!\wedge\del_\xi$, and a right-handed plane $\del_-\!\wedge\del_{\bar\xi}$. These can be realized in terms of null coordinates $(x^+,x^-,\xi,\bar\xi)$, defined as:
\begin{align}
	x^\pm = \frac{x\pm t}{2} \ ; \quad \xi = \frac{z - iy}{2} \ ; \quad  \bar\xi = \frac{z + iy}{2} \ . \label{eq:basis}
\end{align}
The chiral vertices take their simplest form if we redefine the fields $\Phi_h$ through lightlike derivatives:
\begin{align}
	\phi_h \equiv (\del_-)^{-h}\Phi_h \ . \label{eq:phi}
\end{align}
The potentially negative power of $\del_-$ is a normal feature of the lightcone formalism, and doesn't imply non-locality. The significance of the shifted fields \eqref{eq:phi} is that they're closely related to \emph{linearized curvatures}: the electromagnetic field strength for $s=1$, the Weyl tensor for $s=2$, and their HS generalizations. In particular, for non-negative helicities $h=+s$, all the components of the right-handed curvature can be expressed as (see e.g. \cite{Lang:2025rxt}):
\begin{align}
	C_{\dot\alpha_1\dots\dot\alpha_{2h}} = z^{h+1}\,q^{\alpha_1}\dots q^{\alpha_{2h}}\del_{\alpha_1\dot\alpha_1}\dots\del_{\alpha_{2h}\dot\alpha_{2h}}\phi_h \ . \label{eq:C}
\end{align}
Here, the $\alpha$'s and $\dot\alpha$'s are left-handed and right-handed spinor indices, $q^\alpha$ is the spinor square root of the lightlike direction $\del_-$, and we used the AdS warp factor $z$ for concreteness. 

Now, consider three fields $\phi_{h_i}$ ($i=1,2,3$) with helicities $h_i\geq 0$, satisfying the ``triangle inequalities'' $n_i\equiv h_{\text{tot}}-2h_i\geq 0$. One can then construct a cubic vertex by contracting the spinor indices of the three linearized curvatures \eqref{eq:C} (contracting $n_3$ index pairs between the 1st and 2nd curvatures, and so on cyclically). To express the result in terms of the fields $\phi_{h_i}$, let $\del^{(i)}_\mu$ denote a spacetime derivative acting on the $i$'th field. We then denote derivatives along the lightray $\del_-$ and the left-handed plane $q^\alpha q^\beta$ (namely, $\del_-\wedge\del_\xi$) as:
\begin{align}
	\begin{split}
  	\beta_i &\equiv \del_-^{(i)} \ ; \\
	\bbP_{ij} &\equiv q^\alpha q^\beta \del^{(i)}_{\alpha\dot\alpha}\del_\beta^{(j)\dot\alpha} = \del_-^{(i)}\del_\xi^{(j)} - \del_-^{(j)}\del_\xi^{(i)} = -\bbP_{ji} \ . 
	\end{split} \label{eq:derivs}
\end{align}
The cubic vertex in AdS then reads very simply:
\begin{align} 	
	V_{h_1h_2h_3} = \frac{z^{h_{\text{tot}}-1}}{\Gamma(h_{\text{tot}})}\,\bbP_{23}^{n_1}\bbP_{31}^{n_2}\bbP_{12}^{n_3}\,\phi_{h_1}\phi_{h_2}\phi_{h_3} \ , \label{eq:V_chiral}
\end{align}
where the numerical factors $\Gamma(h_{\text{tot}})$ are the ones that will guarantee higher-order consistency of Chiral HS Gravity \cite{Metsaev:1991nb,Metsaev:1991mt,Ponomarev:2016lrm,Skvortsov:2018uru}, as well as agreement with HS Gravity's holographic dual \cite{Kozaki:2025jrj}. The vertices \eqref{eq:V_chiral} are well-known, and are regarded in the literature as the ``easy'' ones. In particular, they're normally understood \emph{not} to include the cubic vertices of Yang-Mills and GR, whose helicity configurations $(+1,+1,-1)$ and $(+2,+2,-2)$ violate both sets of inequalities $h_i\geq 0$ and $n_i\geq 0$. 

The key insight of our paper is this: \emph{the vertex formula \eqref{eq:V_chiral} is valid for all helicities with $h_{\text{tot}}>0$, regardless of the inequalities $h_i\geq 0$ and $n_i\geq 0$ that enable its interpretation in terms of linearized curvatures}. This idea has been overlooked for decades, likely because when $n_i\geq 0$ is violated, the formula \eqref{eq:V_chiral} becomes \emph{non-local}: the spatial derivatives $\del_\xi$, contained in the $\bbP_{ij}$'s, enter with negative powers. However, it's easy to bring \eqref{eq:V_chiral} to a local form: one simply needs to integrate by parts, removing all derivatives from e.g. the 3rd field. The prescription reads: 
\begin{gather}
	\bbP_{23} \to \bbP_{12} - \beta_2\del_z ; \quad \bbP_{31} \to \bbP_{12} + \beta_1\del_z \ , \label{eq:by_parts}
\end{gather}
where the $\del_z$ derivatives are understood to act on the explicit position dependence $z^{h_{\text{tot}}-1}$ in \eqref{eq:V_chiral}. The vertex \eqref{eq:V_chiral} then becomes:
\begin{align}
	   &V^{\text{local}}_{h_1h_2h_3} = \phi_{h_3}\sum_{k_1,k_2=0}^\infty \frac{(-1)^{k_2}}{\Gamma(h_{\text{tot}}-k_1-k_2)}\binom{n_1}{k_2}\binom{n_2}{k_1} \nonumber \\
	     &\qquad \times z^{h_{\text{tot}}-k_1-k_2-1}\,\beta_1^{k_1}\beta_2^{k_2}\,\bbP_{12}^{h_{\text{tot}}-k_1-k_2}\,\phi_{h_1}\phi_{h_2} \ , \label{eq:V_local}
\end{align}
where $\binom{n}{k}$ are the Taylor coefficients of $(1+x)^n$. The series automatically truncates once the argument of $\Gamma(h_{\text{tot}}-k_1-k_2)$ vanishes (i.e. when the $\del_z$ derivatives deplete all the factors of $z$). Thus, only positive powers of $\bbP_{12}$ appear, and locality is restored. While slightly more complicated than the original \eqref{eq:V_chiral}, the local form \eqref{eq:V_local} is still much simpler and more explicit than the previous attempt in the literature \cite{Metsaev:2018xip}. There, the vertices were sought in terms of the symmetric combination $\bbP\equiv\frac{1}{3}(\bbP_{12}+\bbP_{23}+\bbP_{31})$, which didn't readily reveal the simple structure of \eqref{eq:V_chiral},\eqref{eq:V_local}. 

To obtain the flat limit of \eqref{eq:V_chiral}-\eqref{eq:V_local}, one replaces the $z$ factors by a large constant. The $\del_z$ derivatives in \eqref{eq:by_parts} are then negligible, so the $\bbP_{ij}$'s are all equal up to integration by parts. The local formula \eqref{eq:V_local} reduces to just the $k_1=k_2=0$ term, which is the familiar flat vertex $\sim\bbP_{12}^{h_{\text{tot}}}$. The de Sitter vertices can be obtained from the AdS vertices \eqref{eq:V_chiral},\eqref{eq:V_local} by replacing the $z$ factors with $t$, as established in \cite{Kozaki:2025jrj} for vertices constructed out of the derivatives \eqref{eq:derivs}. All the results below can be extended to de Sitter following the analytic-continuation procedure of \cite{Kozaki:2025jrj}. With this understanding, we will focus on the AdS vertices \eqref{eq:V_chiral},\eqref{eq:V_local} and their flat limit. 

We haven't yet proven that the vertex formula \eqref{eq:V_chiral} is correct, i.e. that it's consistent at cubic order for arbitrary helicities with $h_{\text{tot}}>0$. Also, we must prove the claim \eqref{eq:L_full} that the theory with only chiral cubic vertices is consistent to all orders. Those are tasks of the next two sections.

\section{Consistency at cubic order}

Let us derive the cubic consistency of the vertices \eqref{eq:V_chiral}. At this level, each triple of helicities $(h_1,h_2,h_3)$ can be considered separately. In the lightcone formalism of \cite{Metsaev:2018xip}, consistency refers to the closure of the AdS algebra, which we express as the conformal algebra $(P^a,J^{ab},D,K^a)$ of the 3d spacetime $x^a=(t,x,y)$, or equivalently $x^a=(x^+,x^-,y)$. The algebra's generators are constructed out of the fields $\phi_h$ on the $x^+=0$ null hyperplane $(x^-,y,z)$, or equivalently $(x^-,\xi,\bar\xi)$. The fields' canonical commutator reads:
\begin{align}
	 [\phi_h(x),\phi_{h'}(x')] = \frac{(-1)^h\delta_{-h,h'}}{2i\del_-}\,\delta^3(x - x') \ , \label{eq:commutator}
\end{align}
where the delta function is w.r.t. the measure:
\begin{align}
	d^3x \equiv dx^-dydz = -\frac{i}{2}dx^-d\xi d\bar\xi \ . \label{eq:measure}
\end{align}
Each AdS symmetry generator $G$ can be expanded as $G=G_{[2]}+G_{[3]}+\dots$, where $G_{[2]}$ is the free-field quadratic piece, and $G_{[3]}$ are the cubic interacting corrections. For any operator $\calO$ constructed from e.g. three fields $\phi_{h_i}$, the action of a quadratic generator $G_{[2]}$ reduces to some linear differential operator $G_{\text{lin}}$ acting on the $\phi_{h_i}$:
\begin{align}
   \begin{split}
	 &i\left[G_{[2]},\calO(\phi_{h_1},\phi_{h_2},\phi_{h_3})\right] \equiv G_{\text{lin}}\triangleright\calO \\
	 &\quad = \calO\big(G_{\text{lin}}\phi_{h_1},\phi_{h_2},\phi_{h_3}\big) + \calO\big(\phi_{h_1},G_{\text{lin}}\phi_{h_2},\phi_{h_3}\big) \\
	 &\qquad\qquad\qquad + \calO\big(\phi_{h_1},\phi_{h_2},G_{\text{lin}}\phi_{h_3}\big) \ .
   \end{split} \label{eq:linear_action}
\end{align}
For our analysis, we'll need the linear form $G_{\text{lin}}$ of three generators -- the ``Hamiltonian'' $P^-$, the boost $J^{-y}$, and the special conformal generator $K^y$:
\begin{align}
  P^-_{\text{lin}} &= -\frac{\del_y^2 + \del_z^2}{2\del_-} = -\frac{\del_\xi\del_{\bar\xi}}{2\del_-} \ ; \label{eq:P_linear} \\
  J^{-y}_{\text{lin}} &= x^-\del_y - y P^-_{\text{lin}} - ih \frac{\del_{\bar\xi}}{\del_-} \ ; \label{eq:J_linear} \\
  K^y_{\text{lin}} &= 2\xi\bar\xi\del_y - y(x^-\del_- + y\del_y + z\del_z + 1) + 2ih\xi \ . \label{eq:K_y}
\end{align}
There are only 3 generators that receive interacting corrections: $P^-$, $J^{-y}$, and $K^-$. These are called ``dynamical'', while the others, such as $K^y$, are called ``kinematical''. The cubic corrections can all be constructed from the vertex $V_{[3]}$ and the kinematical generator $K^y$ as:
\begin{align}
	P^-_{[3]} &= c\int d^3x\,V_{[3]} \ ; \label{eq:P_cubic} \\
	J^{-y}_{[3]} &= i\left[K^y_{[2]}, P^-_{[3]} \right] = K^y_{\text{lin}}\triangleright P^-_{[3]} \ ; \label{eq:J_cubic_raw} \\
	K^-_{[3]} &= -i\left[K^y_{[2]}, J^{-y}_{[3]}\right] = -K^y_{\text{lin}}\triangleright J^{-y}_{[3]} \ . \label{eq:K_cubic_raw}
\end{align}
Plugging \eqref{eq:K_y}-\eqref{eq:P_cubic} and our vertex formula \eqref{eq:V_chiral} into \eqref{eq:J_cubic_raw}-\eqref{eq:K_cubic_raw}, a straightforward calculation gives (neglecting total derivatives in the $d^3x$ integrals throughout):
\begin{align}
	J^{-y}_{[3]} &= -c\int d^3x\,y V_{[3]} \ ; \label{eq:J_cubic} \\ 
	K^-_{[3]} &= \frac{c}{2}\int d^3x\,(y^2 + z^2) V_{[3]} = 2c\int d^3x\,\xi\bar\xi\,V_{[3]} \ . \label{eq:K_cubic}
\end{align}
This is in fact the simplest possible result, containing only the ``orbital'' terms that follow from the Hamiltonian density $cV_{[3]}$. This is an advantage of the specific vertex formula \eqref{eq:V_chiral}: adding total derivatives, e.g. to reach the local form \eqref{eq:V_local}, generally adds ``intrinsic'' correction terms to \eqref{eq:J_cubic}-\eqref{eq:K_cubic}. Such terms appear in previous treatments \cite{Metsaev:2018xip}, even in the flat limit \cite{Ponomarev:2016lrm}.

Finally, we're ready to check the vertices' consistency at cubic order. We build on the general analysis of \cite{Metsaev:2018xip}, which shows that, for vertices constructed from the derivatives \eqref{eq:derivs}, the only non-trivial commutator to check is $[P^-,J^{-y}]=0$. At cubic order, this requires:
\begin{align}
	P^-_{\text{lin}}\triangleright J^{-y}_{[3]} - J^{-y}_{\text{lin}}\triangleright P^-_{[3]} = 0 \ .
\end{align}
Plugging in \eqref{eq:V_chiral},\eqref{eq:P_linear}-\eqref{eq:J_linear},\eqref{eq:P_cubic},\eqref{eq:J_cubic}, this can indeed be verified by straightforward calculation.
 
\section{Consistency at higher orders}

Let us now prove the higher-order consistency of the chiral theory with just the cubic vertices \eqref{eq:V_total},\eqref{eq:V_chiral}. Again, this requires checking the key commutator $[P^-,J^{-y}]=0$, this time without stopping at cubic order. The only higher order that appears is quartic, where we must demonstrate:
\begin{align}
	\left[P^-_{[3]}, J^{-y}_{[3]} \right] = -c^2\left[\int d^3x\,V_{[3]}, \int d^3x\,yV_{[3]} \right] = 0 \ , \label{eq:quartic_raw}
\end{align}
where in the first equality we used \eqref{eq:P_cubic},\eqref{eq:J_cubic}. Let us express the vertices \eqref{eq:V_chiral} in terms of their flat limit:
\begin{align} 
	V_{h_1h_2h_3} &= z^{h_{\text{tot}}-1}\,V^{\text{flat}}_{h_1h_2h_3} \ ; \label{eq:V_AdS} \\
	V^{\text{flat}}_{h_1h_2h_3} &= \frac{1}{\Gamma(h_{\text{tot}})}\,\bbP_{23}^{n_1}\bbP_{31}^{n_2}\bbP_{12}^{n_3}\,\phi_{h_1}\phi_{h_2}\phi_{h_3} \ . \label{eq:V_flat}
\end{align}
Here, in the flat vertices, we drop the dimensionful constant that should replace the $z$ factors; this won't affect the derivation below. 

Now, note that the vertex \eqref{eq:V_chiral} and canonical commutator \eqref{eq:commutator} contain derivatives only within the left-handed $\del_-\wedge\del_\xi$ plane. Thus, in the $d^3x$ integrals \eqref{eq:measure}, the $d\bar\xi$ integration just factors out, and it's sufficient to focus on the plane $x^+=\bar\xi=0$, where we have $z=\xi$ and $y=i\xi$. Thus, the desired relation \eqref{eq:quartic_raw} takes the form:
\begin{align}
	\left[\int d^2x \sum \xi^{h_{\text{tot}}-1}\,V^{\text{flat}}, \int d^2x\sum \xi^{h_{\text{tot}}}\,V^{\text{flat}} \right] = 0 \ , \label{eq:AdS_quartic}
\end{align}
where the measure is $d^2x \equiv dx^-d\xi$, the sum is over helicity triples with $h_{\text{tot}}>0$, and we drop the helicity labels on the vertices for brevity. Now, we know that the quartic consistency \eqref{eq:quartic_raw} holds \emph{in the flat limit} \cite{Bardeen:1995gk,Siegel:1992wd,Ponomarev:2016lrm,Ponomarev:2017nrr,Serrani:2025owx}, where it reads:
\begin{align} 
	\left[\int d^2x \sum V^{\text{flat}}, \int d^2x\sum \xi\,V^{\text{flat}} \right] = 0 \ . \label{eq:flat_quartic}
\end{align}
We will now demonstrate that the AdS relation \eqref{eq:AdS_quartic} can be derived from its flat counterpart \eqref{eq:flat_quartic}, via a conformal transformation. It isn't obvious that this should work, since the interacting theory (outside of the Yang-Mills sector $h_{\text{tot}}=1$) is \emph{not} conformally invariant. However, it's also not completely surprising, since our vertex formula \eqref{eq:V_chiral} can be traced back to the curvatures \eqref{eq:C}, which \emph{are} conformal.

Consider, then, the conformal generator $K^\xi = \frac{1}{2}(K^z - iK^y)$. Again, it isn't part of the AdS algebra, and thus not a symmetry of the interacting theory. However, it \emph{is} a symmetry of the \emph{free} theory, and thus of the canonical commutator \eqref{eq:commutator}. On the  $x^+=\bar\xi=0$ plane, its linear action (c.f. \eqref{eq:linear_action}) reads simply:
\begin{align}
	K^\xi_{\text{lin}} = -\xi(x^-\del_- + \xi\del_\xi + 1 - 2h) \ .
\end{align}
 A straightforward calculation yields, for any power $m$:
\begin{align}
	K^\xi_{\text{lin}}\triangleright\int d^2x\,\xi^m\,V^{\text{flat}} = (m - h_{\text{tot}})\int d^2x\,\xi^{m+1}\,V^{\text{flat}} \ . \nonumber
\end{align}
Iterating this, we construct the exponentiated action of $K^\xi_{\text{lin}}$ on both sides of the flat consistency relation \eqref{eq:flat_quartic}:
\begin{align}
	e^{-aK^\xi_{\text{lin}}}\triangleright\int d^2x\,V^{\text{flat}} &= \int d^2x\,(1 + a\xi)^{h_{\text{tot}}}\,V^{\text{flat}} \ ; \\
	e^{-aK^\xi_{\text{lin}}}\triangleright\int d^2x\,\xi\,V^{\text{flat}} &= \int d^2x\,\xi(1 + a\xi)^{h_{\text{tot}}-1}\,V^{\text{flat}} \ . \nonumber
\end{align}
Since $K^\xi$ is a symmetry of the commutators, these must still commute with each other as in \eqref{eq:flat_quartic}, for any value of $a$:
\begin{align} 
  \begin{split}
	\bigg[&\int d^2x \sum(1 + a\xi)^{h_{\text{tot}}}\, V^{\text{flat}}, \\
  	  &\int d^2x\sum \xi(1 + a\xi)^{h_{\text{tot}}-1}\,V^{\text{flat}} \bigg] = 0 \ .
  \end{split} \label{eq:transformed_flat}
\end{align}
We now wish to derive the AdS consistency relation \eqref{eq:AdS_quartic} from the conformally-transformed flat relation \eqref{eq:transformed_flat}. To do this, let us isolate the coefficient of a particular quadruple of fields $\prod_{i=1}^4\phi_{h_i}$ in the quartic expressions \eqref{eq:AdS_quartic},\eqref{eq:transformed_flat}. There are 6 contributions to this coefficient, with different triples of helicities inside each factor in the commutator (recall that the first factor in  \eqref{eq:AdS_quartic},\eqref{eq:transformed_flat} comes from $P^-_{[3]}$, and the second -- from $J^{-y}_{[3]}$). We denote these contributions as:
\begin{align}
    	\Big(\{h_1,h_2,-h_5\},\{h_3,h_4,h_5\}\Big) \ ; \ \Big(\{h_3,h_4,h_5\},\{h_1,h_2,-h_5\}\Big) \ ; \label{eq:helicities_1} \\
    	\Big(\{h_3,h_1,-h_5\},\{h_2,h_4,h_5\}\Big) \ ; \ \Big(\{h_2,h_4,h_5\},\{h_3,h_1,-h_5\}\Big) \ ; \label{eq:helicities_2} \\
	    \Big(\{h_2,h_3,-h_5\},\{h_1,h_4,h_5\}\Big) \ ; \ \Big(\{h_1,h_4,h_5\},\{h_2,h_3,-h_5\}\Big) \ , \label{eq:helicities_3}
 \end{align}
where in each case we list the helicities in the two factors. $\pm h_5$ denotes the ``exchanged'' helicity, i.e. the helicity of the fields that are contracted through the fundamental commutator \eqref{eq:commutator}. $h_5$ is summed over, with the sums cut off by the requirement $h_{\text{tot}}>0$ in each triple. Note that the sum of the $h_{\text{tot}}$'s in the two factors is always just the sum of external helicities $\sum_{i=1}^4h_i$. Now, consider the expansion of \eqref{eq:transformed_flat} in powers of $a$. The lowest (i.e. zeroth) power just reproduces the flat consistency relation \eqref{eq:flat_quartic}. The \emph{highest} power of $a$ for given external helicities $h_i$ is naively $\sum_{i=1}^4h_i - 1$, but its coefficients cancel between the two terms in each row of \eqref{eq:helicities_1}-\eqref{eq:helicities_3}. The highest \emph{non-vanishing} power of $a$ in \eqref{eq:transformed_flat} is thus $\sum_{i=1}^4h_i - 2$. Again summing the terms in each row of \eqref{eq:helicities_1}-\eqref{eq:helicities_3}, it's easy to see that the coefficient of this power of $a$ gives the desired AdS consistency relation \eqref{eq:AdS_quartic}. 

This concludes our proof that the AdS chiral theory with only cubic vertices \eqref{eq:V_total},\eqref{eq:V_chiral} inherits all-orders consistency from its flat counterpart. In the Appendix, we supplement this somewhat formal proof with a more hands-on derivation for the simplest non-trivial theory, namely Self-Dual GR.

\section{Outlook}

In this paper, we presented remarkably simple vertices \eqref{eq:V_chiral} and dynamical symmetry generators \eqref{eq:J_cubic}-\eqref{eq:K_cubic} for the cubic interactions of massless fields of any spin in (A)dS in the lightcone formalism. Our vertices are a straightforward, but non-local, generalization of the ``abelian'' ones derived from linearized curvatures. We produced a manifestly local version in \eqref{eq:V_local}. We also showed that any chiral theory that is consistent to all orders in flat spacetime with only cubic vertices, such as Self-Dual GR or Chiral HS Gravity, enjoys the same property in (A)dS. It would be interesting to extend these results to half-integer spins, potentially with supersymmetry. 

We are excited to apply the simple cubic vertices \eqref{eq:V_chiral},\eqref{eq:V_local} to bulk computations in (Chiral) HS Gravity. One application would be to (A)dS/CFT boundary correlators, especially as they pertain to constructing a holographic dual for the chiral theory \cite{Jain:2024bza,Skvortsov:2026ofl}. Another application would be to ``scattering amplitudes'' in the static patch of de Sitter space \cite{Albrychiewicz:2020ruh,Albrychiewicz:2021ndv,Kozaki:2025jrj}. Specifically, the vertex formula \eqref{eq:V_local} can be directly plugged into the kinematical framework of \cite{Kozaki:2025jrj}, and one can start studying the resulting amplitudes for higher-spin symmetry patterns and holographic interpretations.

\begin{acknowledgments}
We are grateful to Julian Lang for discussions, and for double-checking some calculations. We thank Evgeny Skvortsov for email exchanges. We are supported by the Quantum Gravity Unit of the Okinawa Institute of Science and Technology Graduate University (OIST).
\end{acknowledgments}

\begin{widetext}

\appendix
\section{Quartic consistency derivation for Self-Dual GR}

In this Appendix, we present an explicit derivation of higher-order consistency \eqref{eq:quartic_raw} for Self-Dual GR in AdS, without employing a conformal transformation from the flat case. As in the main text, we work on the plane $x^+=\bar\xi=0$, where we can substitute $z=\xi$ and $y=i\xi$, and the relation to prove is \eqref{eq:AdS_quartic}. 

 In Self-Dual GR, the only fields are $\phi_{\pm 2}$, i.e. the positive and negative helicities of the graviton, and the only vertex is the one with helicities $h_i=(+2,+2,-2)$. This vertex has $h_{\text{tot}}=2$ and $n_i=(-2,-2,+6)$. Thus, the desired AdS consistency relation \eqref{eq:AdS_quartic} reads:
 \begin{align}
 	\left[\int d^2x\,\xi\,V^{\text{flat}}, \int d^2x\,\xi^2\,V^{\text{flat}} \right] = 0 \ , \label{eq:GR_commutator}
 \end{align}
where the flat vertex $V^{\text{flat}}$ is given by \eqref{eq:V_flat}. The only contribution to the quartic consistency relation \eqref{eq:GR_commutator} has external helicities $(h_1,h_2,h_3,h_4)=(+2,+2,+2,-2)$. In the notation of \eqref{eq:helicities_1}-\eqref{eq:helicities_3}, the exchanged helicity is then $h_5=+2$.

To evaluate \eqref{eq:GR_commutator}, we must sum over the helicity assignments \eqref{eq:helicities_1}-\eqref{eq:helicities_3}. For each helicity assignment, we apply the fundamental commutator \eqref{eq:commutator} to the contracted fields $\phi_{\pm h_5}$, after removing all derivatives from $\phi_{\pm h_5}$ within each cubic factor $\xi\,V^{\text{flat}}$, $\xi^2\,V^{\text{flat}}$. For this, we integrate by parts, just as in \eqref{eq:by_parts}, except now the explicit coordinate factors are $\xi$ rather than $z$, and we denote the derivatives acting on them as $\del_\xi$. Thus, in a cubic vertex with field labels $(i,j,5)$, the prescription for integrating by parts to remove derivatives from field no. 5 reads:
\begin{align}
	\bbP_{i5} \longrightarrow -\bbP_{ij} - \beta_i\del_\xi \ . \label{eq:by_parts_xi}
\end{align}
Applying this to the cubic factors in the first helicity assignment of \eqref{eq:helicities_1}, we get (omitting the fields factors $\phi_{h_i}$ and overall numerical factors):
\begin{align}
	\xi\,V^{\text{flat}} &\sim \xi\,\bbP_{12}^6\,\bbP_{25}^{-2}\,\bbP_{51}^{-2} &&\longrightarrow&& \left( \xi\,\bbP_{12}^2 + 2(\beta_2-\beta_1)\bbP_{12} \right) \ ; \label{eq:P_1} \\
	\xi^2\,V^{\text{flat}} &\sim \xi^2\,\bbP_{34}^{-2}\,\bbP_{45}^{-2}\,\bbP_{53}^6
	   &&\longrightarrow&& \left( \xi^2\,\bbP_{34}^2 + 4\xi(3\beta_3 + \beta_4)\bbP_{34} + 6(5\beta_3^2 + 4\beta_3\beta_4 + \beta_4^2) \right) \ , \label{eq:J_1}
\end{align}
and similarly for the second helicity assignment of \eqref{eq:helicities_1}:
 \begin{align}
 	\xi\,V^{\text{flat}} &\sim \xi\,\bbP_{34}^{-2}\,\bbP_{45}^{-2}\,\bbP_{53}^6 &&\longrightarrow&& \left( \xi\,\bbP_{34}^2 + 2(3\beta_3+\beta_4)\bbP_{34} \right) \ ; \label{eq:P_2} \\
 	\xi^2\,V^{\text{flat}} &\sim \xi^2\,\bbP_{12}^6\,\bbP_{25}^{-2}\,\bbP_{51}^{-2}\,
 	   &&\longrightarrow&& \left( \xi^2\,\bbP_{12}^2 + 4\xi(\beta_2-\beta_1)\bbP_{12} + 2(3\beta_1^2 - 4\beta_1\beta_2 + 3\beta_2^2) \right) \ . \label{eq:J_2}
 \end{align}
 Note that \eqref{eq:P_1},\eqref{eq:P_2} are nothing but special cases of the local form \eqref{eq:V_local} of the AdS vertex. 
 
 Let us now apply the fundamental commutator \eqref{eq:commutator} to $\phi_{\pm h_5}$. This means multiplying the two lines \eqref{eq:P_1}-\eqref{eq:J_1} along with a factor of $1/(\beta_3+\beta_4)$ (coming from the inverse derivative $1/\del_-$ in \eqref{eq:commutator}), and similarly the two lines \eqref{eq:P_2}-\eqref{eq:J_2} with a factor of $1/(\beta_1+\beta_2)$. Due to momentum conservation $\beta_3+\beta_4 = -(\beta_1+\beta_2)$, some of the terms immediately cancel, and the remainder can be collected as (discarding an overall sign):
 \begin{align}
 	\begin{split}
 	   \frac{2}{\beta_1+\beta_2}\Big( &\xi^2\left\{(3\beta_3+\beta_4)\bbP_{12}^2\bbP_{34} + (\beta_1-\beta_2)\bbP_{12}\bbP_{34}^2 \right\} \\
 	    &+ \xi\left\{ 3(5\beta_3^2+4\beta_3\beta_4 + \beta_4^2)\bbP_{12}^2 - (3\beta_1^2 - 4\beta_1\beta_2 + 3\beta_2^2)\bbP_{34}^2 \right\} \\
 	    &\quad + \left\{6(\beta_2-\beta_1)(5\beta_3^2 + 4\beta_3\beta_4 + \beta_4^2)\bbP_{12} - 2(3\beta_3+\beta_4)(3\beta_1^2 - 4\beta_1\beta_2 + 3\beta_2^2)\bbP_{34} \right\} \Big) \ .
 	\end{split} \label{eq:123_raw}
 \end{align}
 This is the contribution to the desired commutator \eqref{eq:GR_commutator} from the helicity assignments \eqref{eq:helicities_1}. The other helicity assignments \eqref{eq:helicities_2}-\eqref{eq:helicities_3} are given by cyclic permutations on the labels $(1,2,3)$ of the external positive-helicity fields. Thus, to prove the relation \eqref{eq:GR_commutator}, we must show that \eqref{eq:123_raw} vanishes when symmetrized over these permutations. To do this, we manipulate \eqref{eq:123_raw} using the identity:
 \begin{align}
 	\beta_{[i}\bbP_{jk]} = 0 \ , \label{eq:beta_P}
 \end{align}
 along with the integration-by-parts (or momentum conservation) relations:
 \begin{align}
 	\sum_{i=1}^4\beta_i = 0 \ ; \quad \sum_{j\neq i}\bbP_{ij} + \beta_i\del_\xi = 0 \ . \label{eq:by_parts_quartic}
 \end{align}
 We start from higher orders in $\xi$ and work towards the lower ones, picking up lower-order contributions along the way from the $\del_\xi$ derivative in \eqref{eq:by_parts_quartic}. Eventually, we manage to cancel the $\beta_1+\beta_2$ denominator from all the terms in \eqref{eq:123_raw}, bringing \eqref{eq:123_raw} to the form:
 \begin{align}
 	\begin{split}
 	&\xi^2(\bbP_{12} + \bbP_{23} + \bbP_{31}) \bbP_{12}\bbP_{34} \\
 	&\quad + \xi\left\{ (\bbP_{12} + \bbP_{23} + \bbP_{31})\big(5\beta_3\bbP_{12} + 3(\beta_1\bbP_{14} - \beta_2\bbP_{24})\big) + 2(\beta_3\bbP_{12}\bbP_{14} - \beta_2\bbP_{31}\bbP_{34}) \right\} \\
 	&\qquad -(\beta_1+\beta_2+\beta_3)\big(5\beta_3\bbP_{12} + 6(\beta_1\bbP_{14} - \beta_2\bbP_{24})\big) \\ 
 	&\qquad -\bbP_{12}(12\beta_1\beta_2 + 12\beta_2^2) + \bbP_{23}(15\beta_2\beta_3 + 8\beta_1^2 - 12\beta_3\beta_1) - \bbP_{31}(3\beta_3\beta_1 - 12\beta_1^2 + 8\beta_2^2 - 12\beta_1\beta_2) \ ,
 \end{split} \label{eq:123}
 \end{align}
where in some of the terms, we sacrificed manifest symmetry under $1\leftrightarrow 2$ in favor of a shorter expression.
 
Now, when symmetrized over the cyclic permutations of the $(1,2,3)$ labels, \eqref{eq:123} indeed vanishes. The vanishing of the $\sim\xi^2$ term (which corresponds to the flat limit) is due to the identity:
\begin{align}
	\bbP_{[ij}\bbP_{k]l} = 0 \ .
\end{align}
The vanishing upon symmetrization of the $\beta_3\bbP_{12}$ terms in \eqref{eq:123} is due to the identity \eqref{eq:beta_P}. Finally, the $\beta_1\bbP_{14} - \beta_2\bbP_{24}$ and $\beta_3\bbP_{12}\bbP_{14} - \beta_2\bbP_{31}\bbP_{34}$ terms, as well as the last line of \eqref{eq:123}, vanish upon symmetrization manifestly. This concludes our explicit derivation of higher-order consistency for Self-Dual GR, with only the cubic vertex \eqref{eq:V_chiral} with helicities $(+2,+2,-2)$.
 
\end{widetext}

\end{document}